\documentclass[apjl]{emulateapj}
\usepackage{myaasmacros}
\usepackage{mathrsfs}
\usepackage{amsmath}
\usepackage{array}
\usepackage{multirow}
\usepackage{amssymb}
\usepackage{amsfonts}
\usepackage{graphicx}
\usepackage{wasysym}
\usepackage{color}

\newcommand{\gasLow}{{\textit{gasLow}}}
\newcommand{\gasHigh}{{\textit{gasHigh}}}

\newcommand{\pressure}{{\textit{pressure}}}

\newcommand{\Msun}{\ensuremath{\mathrm{M}_\odot }}

\newcommand{\twoFone}[1]{\ensuremath{{}_2F_1}}
\renewcommand{\vec}[1]{\ensuremath{\boldsymbol{#1}}}
\definecolor{grey}{rgb}{0.4,0.6,0.6}
\definecolor{darkgreen}{rgb}{0.0,0.7,0.0}
\definecolor{darkred}{rgb}{0.5,0.,0.}

\def \apjl {{\it ApJ Let.}}
\def \apjs {{\it ApJ Sup.}}

\def \aap {{\it A\&A}}

\shorttitle{Positive Feedback on Disc Galaxies}
\shortauthors{Bieri et al.}
\begin{document}

\title{Playing with positive feedback: external pressure-triggering of a star-forming disc galaxy}

\author{Rebekka Bieri\altaffilmark{1}*,
Yohan Dubois\altaffilmark{1},
Joseph Silk\altaffilmark{1,2,3,4},
Gary A. Mamon\altaffilmark{1}}
\altaffiltext{1}{Institut d'Astrophysique de Paris (UMR 7095: CNRS \& UPMC -- Sorbonne
Universit\'es), 98 bis bd Arago, F-75014 Paris, France}
\altaffiltext{2}{Laboratoire AIM-Paris-Saclay, CEA/DSM/IRFU, CNRS, Univ. Paris
VII, F-91191 Gif-sur-Yvette, France}
\altaffiltext{3}{ Department of Physics and Astronomy, The Johns Hopkins
University Homewood Campus, Baltimore, MD 21218, USA}
\altaffiltext{4}{BIPAC, Department of Physics, University of Oxford, Keble
Road, Oxford OX1 3RH}     
\email{bieri@iap.fr}

\begin{abstract}
Feedback in massive galaxies generally involves quenching of star formation, a
favored candidate being outflows from a central supermassive black hole.  At
high redshifts however, explanation of the huge rates of star formation often
found in galaxies containing AGN may require a more vigorous mode of star
formation than attainable by simply enriching the gas content of galaxies in
the usual gravitationally-driven mode that is associated with the nearby
Universe. Using hydrodynamical simulations, we demonstrate that
AGN-pressure-driven star formation potentially provides the positive feedback
that may be required to generate the accelerated star formation rates observed
in the distant Universe.
\end{abstract}


\maketitle

\section{Introduction}\label{sec:intro} 

The remarkable universality of the Schmidt-Kennicutt star formation law,
ranging from global fits to star-forming galaxies in the nearby Universe
\citep{1998ApJ...498..541K}, local fits to star-forming complexes within
galaxies \citep{Kennicutt+07}, and to star-forming galaxies to $z\sim 2,$
\citep{Genzel+10}  inspires considerable confidence in the theory of
gravitational instability-driven star formation
in galactic disks. There are exceptions, most notably in molecular complexes
with anomalously low star formation rates \citep{Rathborne+14}, but
simple and plausible additions to the usual density threshold criterion for
star formation, most notably by incorporating turbulence, may go far towards
resolving these issues, as demonstrated both theoretically (for a review see
\citealp{2011IAUS..270..179K}) and phenomenologically in well resolved examples
such as  NGC 253 \citep{Leroy+15}.

In the high redshift Universe, the accumulation of recent data on
remarkably  high star formation rates poses a fascinating challenge, 
\citep[e.g.][]{Drouart+14, Piconcelli+15, Rodighiero+15}. Is it
simply a question of turning up the gas fraction or is a new mechanism at work
for inducing more efficient star formation?

In this Letter, we reinforce the case for the latter, more radical view by
simulating the evolution of a fully self-consistent, gas-rich star-forming disk
galaxy that is subject to the overpressuring influence of a vigorous outburst
from its central Active Galactic Nuclei (AGN).  Star formation is a complex
interplay between gas supply, multiphase interstellar medium (ISM), cloud
collapse, and gas ejection from the disk.  We will consider two cases,
corresponding to gas-poor (gas fraction  $10\%$) and gas-rich (gas fraction
$50\%$)  systems at $z\sim 2.$

The case for inducing star formation via AGN activity has been made
analytically \citep{, Silk+Rees1998, Silk+Norman2009} and in simulations
\citep{Gaibler+2012, Ishibashi+Fabian2012,Wagner+2012, Zubovas+2013} with
varying degrees of astrophysical reality.  Simulations demonstrate that whether
the AGN activity is jet or wind-induced is irrelevant: after a few kpc, both
inputs are indistinguishable. Hitherto, however, star formation  in the
multiphase ISM has not been followed in adequate detail because of the lack of
numerical resolution. 

For the first time, we study the effects of pressurization of the disc by
performing simulations in which we fully include self-gravity of the multiphase
ISM and thereby trace the evolution of the star formation rate (SFR) as well as that
of the gas content of the system.


\section{Simulation Set-up}
\label{sec:setup} 

To study the efffect of an external pressure on a galaxy, we have performed a
four isolated disc galaxy simulations with two different initial gas
fractions of 10\%  (hereafter, \gasLow) and 50\% (hereafter,
\gasHigh). 
We allow the galaxies, of one-tenth the total mass of the
Milky Way, to initially adiabatically relax to an
equilibrium configuration (with a reasonable disc thickness) over the rotation
time of the disc at its half-mass radius.  After this first relaxation phase,
we turn on the external pressure, gas cooling, star formation, and also
feedback from supernovae (SNe), as will be described below.

In the initial conditions, the dark matter (DM) particles are sampled with an
NFW~\citep{NFW97} density profile and a concentration parameter of $c=10$ using
the method introduced by \citet{Springel+Hernquist2005}.  For the DM particles,
the virial velocity is set to be $v_{200} = 70 \,\rm km\, s^{-1}$, which
corresponds to a virial radius of $R_{200} \approx 96 \,\rm kpc$ and a virial
mass of $M_{200} \approx 1.1 \times 10^{11}\, \rm \Msun$. We use $10^6$ DM
particles with a mass resolution of $1.21 \times 10^{5}\, \Msun$ to sample the
DM halo.  These stellar disk and bulge were initially sampled with
$5.625\times10^5$ particles of which $6.25\times10^4$ were used
to sample the bulge. The stellar particles are distributed in an exponential
disc with a scale length of $3.44\,\rm kpc$ and scale height $0.2\, \rm kpc$,
and a spherical, non rotating bulge with a Hernquist (\citeyear{Hernquist1990})
profile of scale radius $0.2\, \rm kpc$.  

The simulations are run with the {\sc ramses} adaptive mesh refinement code
\citep{Teyssier2002}. 
The box size is $655\, \rm kpc$ with a coarse level of 7, and a maximum level
of 14 corresponding to a maximum resolution of $\Delta x=40\, \rm pc$. 
The refinement is triggered with a quasi-Lagrangian criterion: if the gas mass
within a cell is larger than $8 \times 10^{7}\, \rm \Msun$ or if more than 8
dark matter particles are within the cell a new refinement level is triggered.

After relaxation, the origin of time was reset to zero and the base simulations
were run further in time with an enhanced and uniform pressure outside the disc
(\pressure\ simulations) for another $\approx 0.42\; \rm Gyr$. This pressure
enhancement is applied instantaneously at $t=0 $ for a value of 3$P_{\rm max}$ 
(hereafter pa3) outside the sphere of radius $r_1 = 12\,\rm kpc$. $P_{\rm max}$ is
the maximum radially averaged pressure in the disk at $t=0$ (reached in the
central few cells), with $P_{\rm max} \simeq 9.8\times 10^{-13}\,\rm Pa$ for
the \gasLow\ simulation set and $\simeq 4.7 \times 10^{-12}\,\rm Pa$ for the
\gasHigh\ simulation set. 

Sub-grid models for cooling \citet{Sutherland+Dopita1993} and star formation,
as well as 
SN feedback, were used in
the simulations.
Gas is turned into star particles in dense cold
regions of gas density $n_{\rm gas} > n_0 = 14\,\rm H \,cm^{-3}$
by drawing a probability from the Schmidt law  $\dot \rho_*=0.01\rho_{\rm gas}/t_{\rm ff}$
to form a star 
with a stellar mass of $m_*=n_0\Delta x^3\simeq2\times 10^4 \,\rm M_\odot$
\citep{Rasera+Teyssier2006}. 
The gas pressure and density are evolved using the Euler equations, with an equation of state  for a
mono-atomic gas with $\gamma=5/3$.
In order to prevent catastrophic and artificial collapse of the
self-gravitating gas, we use a polytropic equation of state $T=T_0(n_{\rm
gas}/n_0)^{\kappa-1}$ to artificially enhance the gas temperature in high gas
density regions ($n_{\rm gas}>n_0$).  Here $\kappa=2$ is the polytropic index,
and $T_0=270\, \rm K$,
chosen to resolve the Jeans length with minimum 4 cells~\citep{Dubois&Teyssier2008}. 
We account for the mass and energy release
from type~II SNe.  The energy injection, which is purely thermal, corresponds
to  
$E_{\rm SN} = \eta_{\text{SN}}\, \left (m_* / {\rm M_\odot} \right) 10^{50} \; \text{erg}$,
where $\eta_{\text{SN}} =0.2$ is the mass fraction of stars going into SNe. 
We also return an amount $\eta_{SN}\,m_*$ back
into the gas for each SN explosion which occurs $10\,\rm Myr$ after the birth
of the star particle.  To avoid excessive cooling of the gas due to our
inability to capture the different phases of the SN bubble expansion, we used
the delayed cooling approach introduced by \citet{Teyssier+2013}.

\section{Results}
\label{sec:results}
\begin{figure*}
 \centering
 \leavevmode
 \includegraphics[width=0.93\hsize]{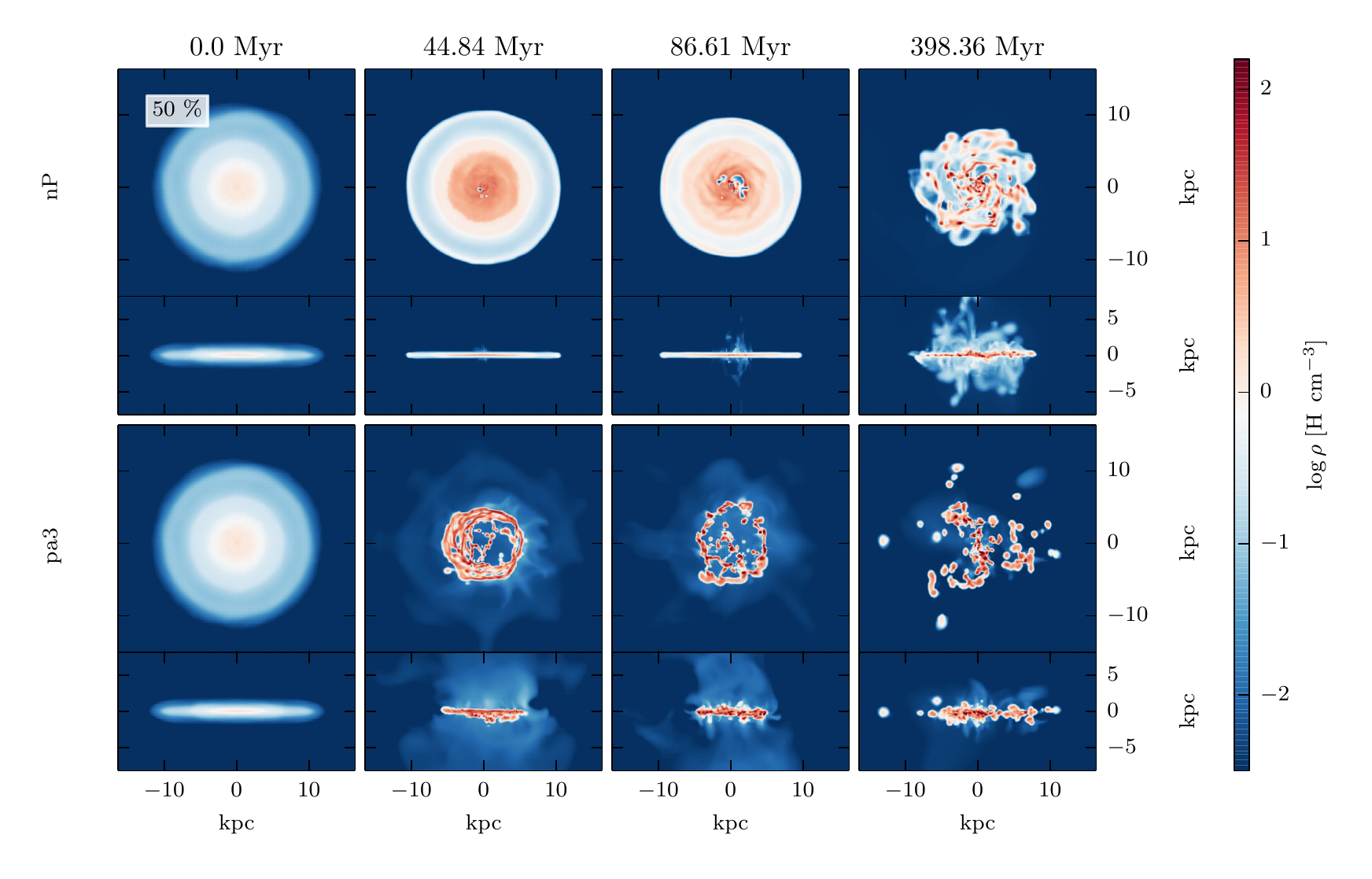}
\caption{Gas density maps (mass-weighted) for the \gasLow\ 
simulations, for no pressure enhancement (top), and for 
a pressure enhancement of a factor 3 (bottom), with time evolving from left
to right. The galaxies are shown both face-on (upper portion of panels) and
edge-on (bottom portion of panels).
}
\label{fig:MapGasLow}
\end{figure*}
\begin{figure*}
 \centering
 \leavevmode
 \includegraphics[width=0.93\hsize]{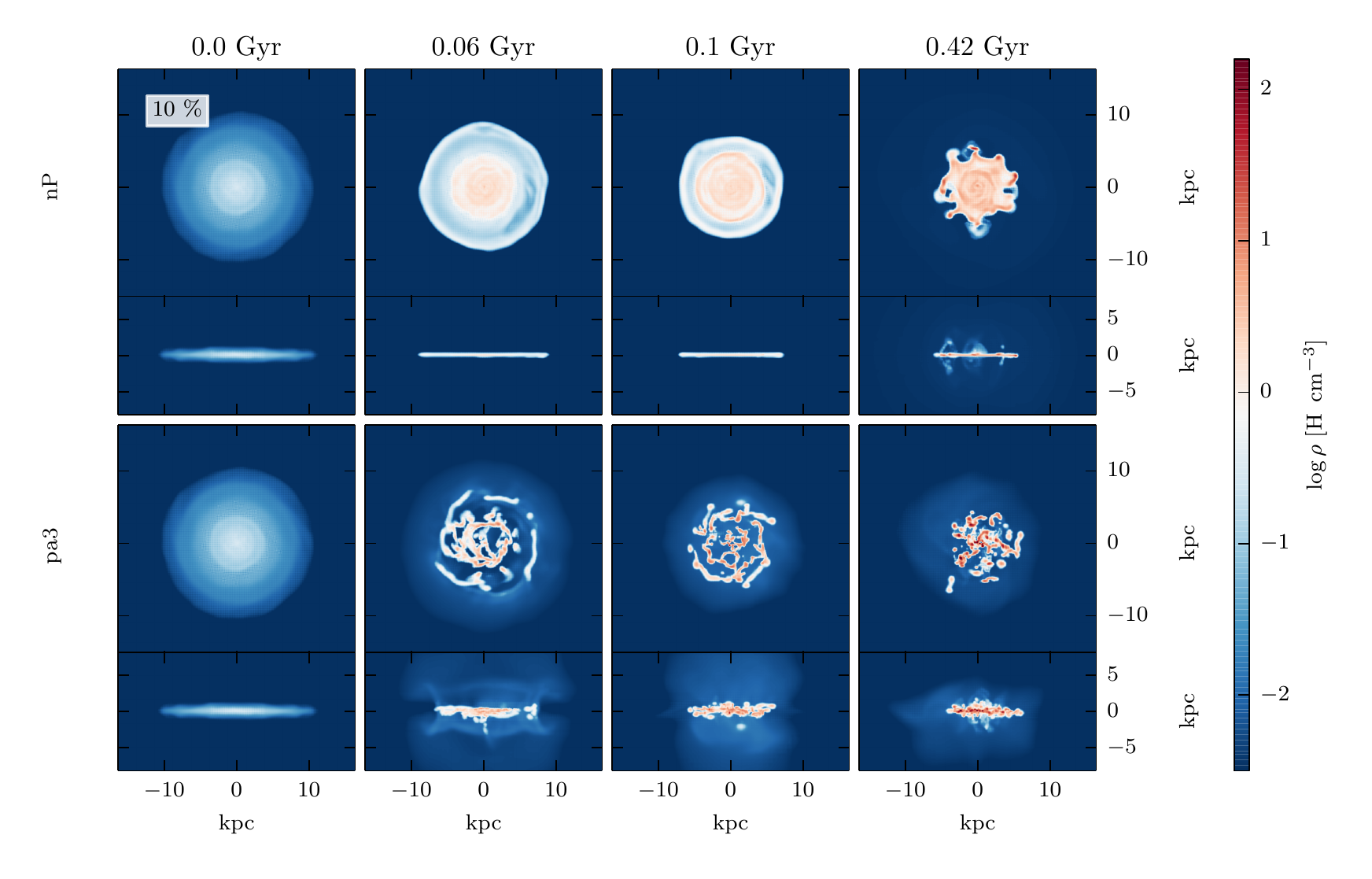}
\caption{Same as in Fig.~\ref{fig:MapGasLow} for the \gasHigh\ simulations. 
}
\label{fig:MapGasHigh}
\end{figure*}
\subsection{Disc fragmentation}

The application of external  pressure at $t=0$ leads to fragmentation of the
gaseous galaxy discs, i.e.  accelerated clump formation.  This can be seen
in Fig.~\ref{fig:MapGasLow} for the \gasLow\  and in
Fig.~\ref{fig:MapGasHigh} for the \gasHigh\ simulations. 
However, the \gasHigh\ case shows more gas between clumps in the enhanced
pressure run, as well as more gas ejection
from the disk.

\begin{figure}
\plotone{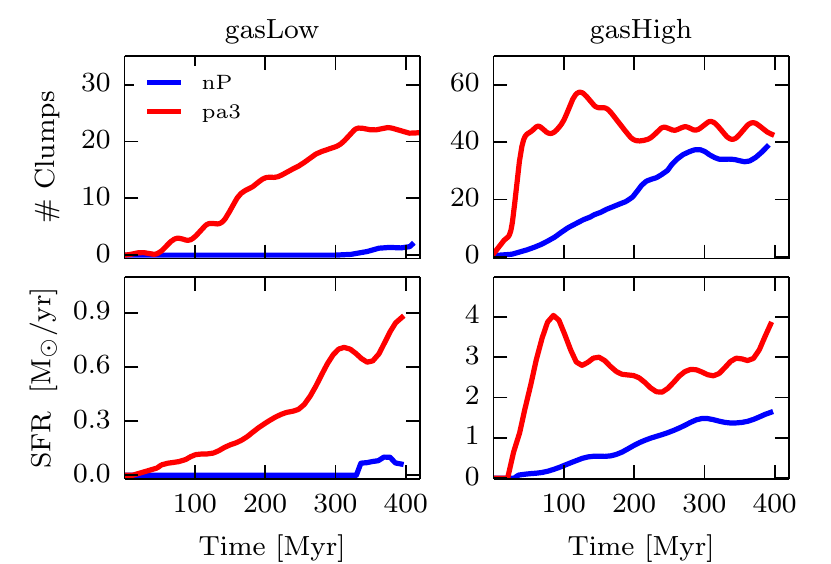}
\caption{Top: Time evolution of the number of clumps.  Bottom: SFR
 as a function of time.  The left panel shows the \gasLow\ simulation
set and the right panel shows the \gasHigh\ simulation set. Here blue denotes
simulations with and green simulations without external pressure applied at $t=0$.} 
\label{fig:CF-SFR}
\end{figure}
\begin{figure}
\plotone{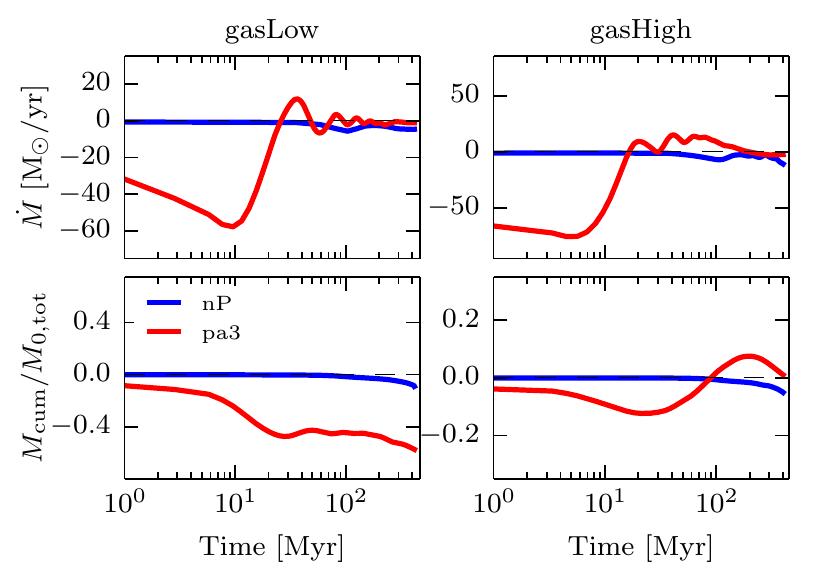}
\caption{Time evolution of the net mass flow rate (top) and cumulative net mass
flow relative to the initial mass $M_{\rm 0,init}$ measured within a galaxy
radius of 12~kpc and a disc height of 1.5~kpc before enhancing the pressure
(bottom), both at 16~kpc from the galaxy center for the \gasLow\ (left), and
\gasHigh\ (right) simulations.}
\label{fig:MassOutflow}
\end{figure} 

Since the star formation recipe depends on the local gas density, we expect an
enhanced star formation when more clumps are formed (gas gets more
concentrated), assuming that the clumps have sufficient mass. Therefore, if 
external pressure leads to increased fragmentation and hence increased clump
formation, we expect the star formation to be positively enhanced when external
pressure is applied on the galaxy.  We first consider the fragmentation by
counting the high-density clumps.   
To find the clumps, we run the clump finder described in
\citet{Bleuler+Teyssier2014}, which identifies all peaks and their highest
saddle points above a given threshold ($n_{\rm gas} > 20\,\rm H\, cm^{-3}$).
A clump is recognized as an individual one when the peak-to-saddle ratio is
above $1.5$, otherwise the peak is merged with the neighbor with which it
shares the highest saddle point.  

The top panel of Fig.~\ref{fig:CF-SFR} shows the number of clumps as
a function of time.  One can see that generally more clumps are formed when
external pressure is applied. In the \gasLow\ simulations, the number of clumps
starts rising soon after this pressure is applied. In the absence of external
pressure, the number of clumps is unity (the disc as a whole corresponds
to a clump). This value is only reached at late times (after 300~Myr) once the
gas has sufficiently collapsed to reach the clump gas density threshold.  

On
the other hand, in the \gasHigh\ simulations,  clump formation 
sets in at earlier times, even without any forcing by external pressure.
This different behavior is related to differences in the \cite{Toomre64} $Q$
parameter in the galaxies of the two simulations, such
that the gaseous disc is stable in the \gasLow\ case 
($\langle Q \rangle = 3.29 > 1$ measured at $t=0$),
but unstable in the \gasHigh\ case ($\langle Q \rangle = 0.72 < 1$).
This demonstrates
that the fragmentation of the disc can be driven by the forcing of an external
pressure, even though the disc is initially Toomre stable.  For the \gasHigh\
simulations, the rise in number of clumps is, however, much faster when
external pressure is applied. At later times, the number of clumps flattens
after $\sim$ 
100~Myr where for the no pressure simulation the number of clumps keeps rising
with time.

\subsection{Mass Flow Rate}\label{subsec:MassFlow} 
The star formation rate is sensitive to both the mass flux and to the clump mass
distribution.  We measure the gas mass flux through a sphere of radius 16~kpc
as
\begin{equation} 
\dot{M} = \oiint \rho \,\vec{v} \cdot \hat{\vec{r}}\, \mathrm{d}S 
= \sum_{i \in \mathrm{shell}} m_{i}\, \vec{v}_{i} \cdot \hat{\vec{r}}_i
\,{S\over V}
\ ,
\label{massflux}
\end{equation} 
where $i$ denotes the index of a cell within a shell of surface $S$ and volume
$V$.  In all simulations with enhanced pressure, an incoming pressure-driven
mass inflow is created at the beginning of the simulation and is followed by a
short mass outflow (seen in the edge-on view of the galaxies in
Figs.~\ref{fig:MapGasLow} and \ref{fig:MapGasHigh}), after which it oscillates
around zero for the remaining of the simulation (top panels of
Fig.~\ref{fig:MassOutflow}).  This is to be expected, as the pressure wave
coming into the galaxy carries momentum causing the galaxy to expel more gas
when compared with the non-pressurized case where the mass flow rate  is close to
zero.

The bottom panels of Fig.~\ref{fig:MassOutflow} shows that the cumulative mass
flow remains negative for the \gasLow\ pressurized simulations, indicating a
greater mass inflow than outflow for these simulations, while the cumulative
mass flow is close to zero for the non-pressurized simulations.  Given that the
SFR enhancement is larger (factor $\sim$ 10) than the net gas change (factor
$\sim$ 2), we argue that the measured mass inflow does not have a major
influence on SFR enhancement measured in the pressurized simulation of \gasLow. 
For the \gasHigh\ simulations, the difference in the cumulative mass flow is
even smaller, showing that the two galaxies have the same amount of gas to form
stars. 

\subsection{Star formation history}
We have seen that an increased pressure enhancement leads to an increased
number of clumps. Since the star formation is proportional to the local density
and since the gas density threshold of clump detection was set to be above that
for star formation, one expects a similar behavior for the SFR.
 
The SFR of the different runs, shown in the bottom panel of
Fig.~\ref{fig:CF-SFR} for the \gasLow\ (left) and \gasHigh\ (right)
simulations, resemble the time evolution of the number of clumps previously
shown in the top panel of Fig.~\ref{fig:CF-SFR}. In particular, for the
\gasLow\ galaxies, while star formation sets in very late at a modest rate in
the simulation without external pressure, the SFR rises almost immediately and
gradually in the simulation  when pressure is initially applied.  For the
\gasHigh\ galaxies, the SFR  rises gradually starting  at early times when no
pressure is enforced on the galaxy, and later accelerates, similarly to the
clump behavior.  In contrast, when external pressure is initially applied to
the \gasHigh\ galaxy, the SFR rises significantly faster at early times,
reaching a plateau after $\sim$80~Myr, maintaining this rate for the remainder
of the simulation. 

\section{Conclusions}\label{sec:conclusions} %

We show that a toy model for AGN-induced overpressurization leads to enhanced
star formation in  disk galaxies.  The effects are dramatic at early times,
regardless of the initial gas fraction of the galaxy, although the high gas
fraction galaxy, which is Toomre unstable, experiences a very early rapid burst
of star formation, followed by more moderate SFR that remains above the SFR in
the non-pressurized case.  However, even in the Toomre stable low gas fraction
galaxy, external pressure can enforce fragmentation and star formation, albeit
on a longer time scale. 

One reason for the increased star formation in pressurized galaxies  is early
pressure-driven mass inflow from the halo and outer disc, on a 10 Myr
time-scale  (Fig.  \ref{fig:MassOutflow}), especially on the low gas fraction
galaxy.  This gas inflow feeds cloud/clump growth in the inner parts of the
galaxy (Fig.  \ref{fig:CF-SFR}) and eventually leads to enhanced star
formation.  The infall effects are more dramatic for the high gas fraction
galaxy, where the disc is more unstable.  (Fig.~\ref{fig:MapGasHigh}).  We have
confirmed this with a detailed analysis of the Toomre parameter and clump
properties, which we defer to a more future study. 

One expects that gas-rich star-forming galaxies should have
gravitationally-driven star formation rates of the order of $100~\Msun \mathrm{
yr}^{-1}$, simply by scaling the gas surface density, and maintaining a similar
efficiency. Indeed, observations at $z\sim 2$ confirm this trend.  The more
exotic cases of very high ($\sim 1000~\Msun \mathrm{yr}^{-1}$) star formation
rates are more typically found at higher redshift and almost invariably have
associated luminous AGN.  While any connection is speculative, and indeed the
very direction of possible causality is vigorously debated, our simple tests of
the effects of AGN-induced pressure, due to wind-driven or jet-initiated bow
shocks that overpressurize the entire inner gaseous disk, suggest that strongly
enhanced star formation rates  are readily achievable.

Similar SFR enhancements are found from increased pressure in the initial
stages of ram pressure harassment, although gas loss dominates the long term
behavior \citep{1999ApJ...516..619F, 1999MNRAS.308..947A, 2014MNRAS.438..444B}.
Here, however, the gas is retained, and the star formation rate enhancement is
far more pronounced.

\section*{Acknowledgments}
This work has been done within the Labex ILP (reference ANR-10-LABX-63) part of
the Idex SUPER, and received financial state aid managed by the Agence
Nationale de la Recherche, as part of the programme Investissements d'avenir
under the reference ANR-11-IDEX-0004-02.  It also has been partially supported
by grant Spin(e) ANR-13-BS05- 0005 of the French ANR.  YD and JS acknowledge
support from ERC project 267117 (DARK) hosted by UPMC -- Sorbonne Universit\'es
and JS for support at JHU by National Science Foundation grant OIA-1124403 and
by the Templeton Foundation.  RB has been supported in part by the Balzan
foundation.  The simulations have made use of the Horizon cluster, for which we
specially thank Stephane Rouberol for technical support.  We also thank M. D.
Lehnert, V. Gaibler, and J. Coles for valuable discussions.

\bibliographystyle{mn2e}
\bibliography{refs}
\end{document}